\documentclass{iopconfser}

\usepackage{natbib}

\input epsf
\def\plotone#1{\centering \leavevmode
\epsfxsize=1.00\columnwidth \epsfbox{#1}}
\def\plotonemedium#1{\centering \leavevmode
\epsfxsize=0.70\columnwidth \epsfbox{#1}}
\def\plotonesmall#1{\centering \leavevmode
\epsfxsize=0.50\columnwidth \epsfbox{#1}}

\begin{document}

\title{Observations of Holographic Quantum-Foam Blurring}

\author{Eric Steinbring}

\affil{National Research Council Canada, Herzberg Astronomy and Astrophysics; 5017 West Saanich Road, Victoria, British Columiba, V9E 2E7, Canada}

\email{Eric.Steinbring@nrc-cnrc.gc.ca}

\begin{abstract}
The ``foamy" nature of spacetime at the Planck scale was an idea first introduced by John Wheeler in the 1950s. And for the last twenty years or so it has been debated whether those inherent uncertainties in time and path-length might also accumulate in transiting electromagnetic wavefronts, resulting in measurable blurring for images of distant galaxies and quasars. A confusing aspect is that ``pointlike" objects will always be blurred out somewhat by the optics of a telescope, especially in the optical. But it turns out that Gamma-Ray Bursts (GRBs) are more useful to test this, and have been observed by a host of ground-based and space-based telescopes, including by the {\it Fermi} observatory for well over a decade. And a recent one was unprecedented: GRB221009A was extremely bright, allowing follow-up from the infrared through the ultraviolet to X-rays and gamma-rays, including a first association with photons at high TeV energies. I will discuss how that observation is in direct tension with the calculus of how spacetime “foaminess” can add up in an image of a pointsource at cosmological distances, which at high-enough energy could spread these out over the whole sky without resulting in photon loss. A simple multiwavelength average of foam-induced blurring consistent with holographic quantum gravity is described, analogous to atmospheric seeing from the ground. This fits with measured instrumental point-spread functions and with the highest-energy localization of GRB221009A, resolving the observational issues and pointing to a key physical implication: spacetime does not look smooth.
\end{abstract}

\section{Introduction}

Although general relativity (GR) is fundamentally smooth, a successful theory of quantum gravity (QG) must incorporate metric graininess.  As energies, lengths and timescales approach the Planck scale it is expected that an inherently ``foamy" microscopic spacetime structure will be revealed, as first proposed by Wheeler \citep[1957; see ][for a review]{Carlip2023}. One phenomenology with which to probe this regime could be whether light propagation from a source some distance $L$ away is measurably affected, for example by violating Lorentz invariance, and giving an energy dispersion $\delta E$ scaling as $L/c$ \citep[as reviewed in][]{Amelino-Camelia2013}. Another would be if images of those sources were blurred: the accumulation of tiny distance fluctuations $\pm \delta l$ proportional to Planck length $l_{\rm P} \sim {10}^{-35}$ m (or equivalently, timescale $t_{\rm P}\sim 10^{-44}~{\rm s}$) in electromagnetic wavefronts, as they are randomly ``kicked" while travelling through the spacetime foam. If so, the amount of phase degradation at observed wavelength $\lambda$ should depend on the summation of small phase perturbations $\Delta \phi = 2\pi \delta l/\lambda$ along their long trajectory of length $L$ as $$\Delta\phi_0=2\pi a_0 {l_{\rm P}^{\alpha}\over{\lambda}}L^{1 - \alpha}\eqno(1)$$for $a_0\sim 1$ and $\alpha$ specifying the QG model: $1/2$ implies a random walk, and would be a strong effect; $2/3$ is weaker, consistent with the holographic principle; and no degradation at all occurs for a value of $\alpha=1$ \citep{Ng2003}. A downside of this latter observational approach is that the impact on distant pointlike objects in optical light might be comparable only to the diffraction-limit of the {\it Hubble Space Telescope} (HST). So, if present it was already known to be weak ($\alpha\leq 0.65$) via images of distant galaxies \citep{Lieu2003, Ragazzoni2003} and active-galactic nucleii \citep[AGNs;][]{Steinbring2007, Christiansen2011, Perlman2011, Tamburini2011}. This report highlights more recent observations of gamma-ray bursts (GRBs) both in the optical and at shorter wavelengths. Those do exhibit blurring consistent with equation 1 (and $\alpha=0.667$) at high energy when observational effects are accounted for. A classical analogy akin to atmospheric scattering helps explain how, for the first time, that shows spacetime is observably not smooth. 

\subsection{Why Gamma-Ray Bursts Provide Better Targets with Which to Sense Blurring}

As the foam-induced blurring effect should grow with shorter observed wavelengths, GRBs provide better targets for probing the QG regime. Beyond providing higher-energy photons, they are near-pointlike, with emission regions known to be less than parsecs across, much more compact than galaxy scales ($\sim 250~{\rm kpc}$). This can overcome the instrumental point-spread function (PSF) limitations of such telescopes, which are significantly larger than diffraction, that is $\lambda/D<<1$, where $D$ is the telescope diameter, typically $\sim 1~{\rm m}$. But it also immediately reveals a tension with equation 1, because for $\alpha=0.667$ blurring should then be obvious in a fairly ``nearby" GRB, even one with redshift of only $z=0.10$, emitting photons at 100 MeV, as each wavefront phase dispersion $\Delta \phi_0$ would be over 1 radian. In the simplest interpretation, that angle is directly equivalent to a deflected ray, ``scattering" photons over nearly the whole sky. And if each photon were so scattered, GRBs could not be identifiable with a galaxy host unless $\alpha$ is larger, as was considered by \cite{Perlman2015}. \cite{Steinbring2015} provides an alternate solution, recognizing that if blurring is a random process, it can be either more or less than equation 1, but then the mean angle-of-arrival of detected photons or the accumulated long-exposure image generated from those must instead include only those photons scattered less than the solid angle visible to the telescope. The cumulative PSF (i.e. for $\Delta\phi\leq 2\pi$) can still, however, include those from along the path to the source redshift, any within the field of view (FoV) in addition to diffraction, and down through to an absolute minimum deflection attributable to the Planck scale. Averaged this way, and designated $\Phi$, this predicts GRBs will be identifiable well beyond 10 GeV and that the general effect, although weaker, is a broadened PSF core falling within a surrounding halo \citep{Steinbring2016}. This is distinct from the interpretation that a larger $\alpha$-value would cause no foam-scattered photons to be detectable outside instrumental diffraction pattern, as in the model of \cite{Ng2022}.

\begin{figure}
\plotone{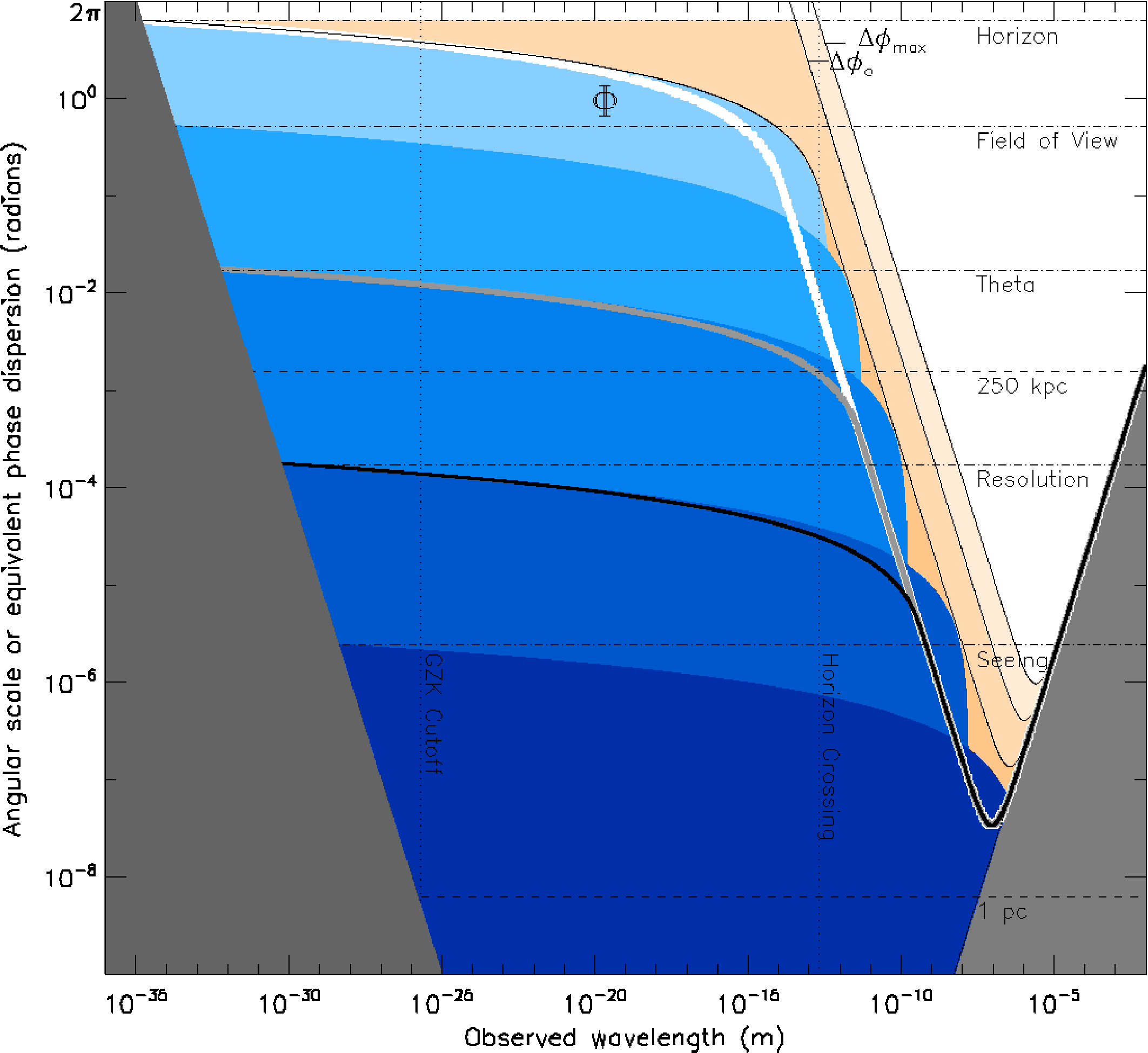}
\caption{Expected angular localization (or equivalent wavefront phase dispersion) of GRB sources on the sky. Maximal limit imposed by foam-induced blurring (red-shaded region; $\alpha=0.650$, $z=4.61$), the lesser limit of equation 1, and still-lower from that due to restricted instrument viewing angles, as defined by equation 4 (blue) are shown; their resulting average PSF shape ($\Phi$) where photons fall inside either the FoV of an all-sky viewing telescope (white curve, $\alpha=0.667$, $z=1.41$) and within characteristic angle Theta (thick grey) or the sharpest instrumental resolution limit (black) are shown; apparent source sizes (dashed horizontal lines, calculated for $z=0.151$) and localization/resolution (horizontal dot-dashed) including typical ground-based seeing (1 arcsec). Vertical dotted lines indicate the wavelength where photons can maximally be scattered to the horizon and, for reference, the Greisen–Zatsepin–Kuzmin cutoff. Dark grey shading indicates unsensed regions due either to diffraction (shown here assuming a telescope of a circular diameter of $D=8$ metres), or at the highest energies, fundamentally limited as wavelengths approach the Planck scale. The lower-resolution limits of realistic $\gamma$-ray telescopes (not diffraction limited) will be addressed in the next sub-section.}\label{figure_model}
\end{figure}

\subsection{Accounting for the Strongest-Possible, Limiting Case of Blurring}

The approach followed here is to start by finding the reddest wavelength with the worst-possible GRB localization. This is equivalent to when a foam-affected photon can be scattered as much as $2\pi$, and so is not localizable because, effectively, that photon could have arrived at the telescope from any angle on the sky. At most, for a wavefront propagating through spacetime foam, the accumulated blurring can be $$\Delta\phi_{\rm max} = \Delta\phi_{\rm los} + \Delta \phi_z = 2\pi a_0 {l_{\rm P}^{\alpha}\over{\lambda}}\Big{\{}\int_0^z L^{1 - \alpha} {\rm d}z + {{(1-\alpha)c}\over{H_0 q_0}}\times \int_0^z (1+z) L^{-\alpha} \Big{[}1 - {{1 - q_0}\over{\sqrt{1 + 2 q_0 z}}}\Big{]} {\rm d}z \Big{\}}$$ $$ = (1+z)\Delta\phi_0, ~~~~~~~~~~~~~~~~~~~~~~~~~~~~~~~~~~~~~~~~~~~~~~\eqno(2)$$where $L = ({c/{H_0 q_0^2}})[q_0 z - (1 - q_0)(\sqrt{1 + 2 q_0 z} - 1)]/(1+z)$ is the comoving distance and deceleration parameter $q_0={{\Omega_0}/{2}} - {{\Lambda c^2}/{3 H_0^2}}$ assumes a standard $\Lambda$CDM cosmology; values of $\Omega_\Lambda=0.7$, $\Omega_{\rm M}=0.3$ and $H_0=70~{\rm km}~{\rm s}^{-1}~{\rm Mpc}^{-1}$ will be used throughout, as in \cite{Steinbring2007}. That effect gets stronger with bluer light; here, $\Delta\phi_{\rm los}$ includes waves propagating from points along the line of sight, and $\Delta \phi_z$ are exclusively those redshifted to the observer. Notice that the ratio between the greatest and the least-possible effect is always $\Delta \phi_{\rm max} / \Delta \phi_{\rm P}=(1 + z) a_0 (L/l_{\rm P})^{1 - \alpha}$, without dependence on $\lambda$, and where $\Delta \phi_{\rm P} = 2\pi{{l_{\rm P}}\over{\lambda}}$, that is, a minimal perturbation corresponding to the Planck length. And so, as long as no photon is scattered to the horizon, a long exposure produces an image averaging all the detectable phase dispersions, and if those have a distribution with amplitude ${\Delta \phi}~\sigma (\Delta \phi) = 1-A \log({{\Delta \phi}/{\Delta \phi_{\rm P}}})$, this is $${1\over{A}} \int \Delta \phi ~\sigma (\Delta \phi) ~{\rm d}{\Delta \phi} = (1 + z) \Delta \phi_0, \eqno(3)$$for $A = 1/\log{[(1 + z) a_0 (L/l_{\rm P})^{1 - \alpha}]}$, recovering equation 1, and constant for all $\lambda$. To put this statement another way: some photons are dispersed by $\Delta\phi_0$ (and perhaps right up to the maximum), but the average blurring at any wavelength is always less, and redward of this ``horizon-crossing'' (vertical dotted line in Figure~\ref{figure_model}) it should maintain a simple power-law wavelength dependence.

\subsection{How Real Telescopes Should Sense Pointsources Blurred by Quantum Foam}

Now imagine what happens to photons blueward of the horizon-crossing wavelength. To do so, it is helpful to consider a point source viewed by a telescope sensitive only to dispersions smaller than some opening angle $\theta$, that is, something less than its FoV. In Figure~\ref{figure_model}, a characteristic angle ``Theta" of $1^\circ$ is chosen to illustrate this. For $\theta \leq 2\pi$ and $A>0$, this implies a PSF mean width \citep{Steinbring2015} $$\Phi = \Phi_R + \Phi_\theta = R\Big{(}{{\lambda}\over{D}}\Big{)}^\rho + \int_0^\theta \Delta \phi ~\sigma (\Delta \phi) ~{\rm d}{\Delta \phi} ~~~~~~~~~~~~~~~~~~~~$$ $$~~~~~~~~~~~~~~~~~~~~~~~~~~~~~~~~~~~~~~~~~ = A R \Big{(}{\lambda\over{D}}\Big{)}^\rho \Big{[} 1 + \log{\Big{(}{{2\pi l_{\rm P} D^\rho}\over{R \lambda^{\rho+1}}}\Big{)}}\Big{]} + \theta \Big{\{} 1 + A\Big{[} 1 + \log{\Big{(}{{2\pi l_{\rm P}}\over{\theta \lambda}}\Big{)}}\Big{]}\Big{\}}, \eqno(4)$$where the right-hand-side involves integration by parts, that is, splitting the integral over $\Delta\phi$ above and below $R(\lambda/D)^\rho$. For a telescope of diameter $D$, the {$\Phi_{\rm R}$} portion includes all phase dispersions up to the instrumental resolution limit, e.g., $R=1.22$ and $\rho=1$ for a perfect, circular and unobstrcted optical telescope, set by diffraction. There is therefore a range of wavelengths where $\Phi$ has to be more than the telescope resolution and less than the FoV. And so, divided by each successive outer scale which is visible to the telescope, there occurs four distinct but self-similar levels of blurring. The narrowest is a sharp PSF core for those photons always falling inside the instrument resolution, and there are three progressively broader layers for those falling either outside Theta or its FoV, and eventually the horizon itself. The last corresponds to the broadest possible blurring, and the sub-regimes are illustrated in Figure~\ref{figure_model} by shading them darker blue with each successively narrower field, for $\alpha=0.650$ and $z=4.61$ (set by previously observed limits, and available data; the reasons for those choices will be made clear in the next section). A fifth level would be due to the natural seeing disk from the ground, which is about 1 arcsec across or less from the best sites. And analogous to quantum foam, this spread occurs due to fluctuations in index-of-refraction in turbulent air along the telescope beam. It is not actually observable because the associated average (foam-induced) effect at optical/near-infrared wavelengths must fall well below diffraction. But the same limiting behaviour, with a similar asymptote towards shorter wavelength, illustrates this interesting comparison. Notice also that these layers all ``turn over'' in the same sense toward higher energy, implying a smoothly-scalable transition between them: at the top, limited by the horizon, and at the lower edge by instrument resolution, and in the middle is Theta.

\subsection{Drawing an Analogy Between Foam-Induced Blurring and a Classical Scattering Process}

Perhaps another helpful classical analogy to think of here, like how incremental wavefront phase-scattering is accumulated in passage through Earth's turbulent atmosphere, may be the game ``Plinko."  In this familiar game of chance, a disk slides down a steep inclined plane, interrupted by uniformly spaced pegs. That pegboard progressively scatters the path of the disk. And clearly, in the final distribution of outcomes along the lower edge of the board, the size of that disk matters when compared to the peg-to-peg distance - especially when those lengthscales are comparable. This should be irrespective of other details such as the peg cross-section (or any reasonable physical material choices for disks and pegs, i.e. whether made of plastic, wood, steel, etc.) and the boundary conditions necessarily imposed from a finite width to the board, and its length. However contrived in reality though, one intuitively expects that when releasing disks in the middle at the top, not every result can be a ``winner" and lie exactly below (for even the luckiest player), instead resulting in a broad distribution clustered in the middle of the board at the bottom. But how wide is the distribution?  In this analogy, Theta is the characteristic distribution width: shortward of the horizon-crossing wavelength we expect to find most (but not all) high-energy photons within that angle relative to the best-resolved (optical) position. To emphasize: a wide-tailed large-Theta distribution means that instruments unable to resolve angles smaller than Theta should still see photons, although those may appear increasingly near-uniform on the sky towards highest energy. This signature behaviour - inducing a broad halo - is obviously distinct from the expectation of finding no photons scattered outside the unresolved cores of pointsources, and is what will be looked for in the GRB data.

\section{Observations and Discussion}

The available GRB data can now be compared to the expected quantum-foam-induced PSF (labelled $\Phi$; here given in equation 4). And a recent detection is special, as it is the brightest, highest-energy one ever observed: GRB221009A. In the discussion to follow, all of the presented data are already summarized in \cite{Steinbring2023}, and are only reproduced here. The intent is to emphasize, particularly with {\it Fermi} LAT, the remarkable adherence to the characteristic PSF and halo behaviour described above. The argument made is that previous studies have not correctly accounted for instrumental effects on the foam-induced PSF, which is why those do not report its detection.

\subsection{The Special Case of GRB221009A in Context with Other Data}

\begin{figure}
\hspace{2 cm}\plotonesmall{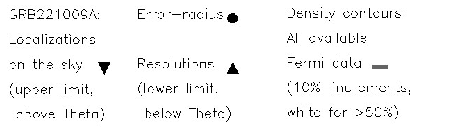}\\
\plotone{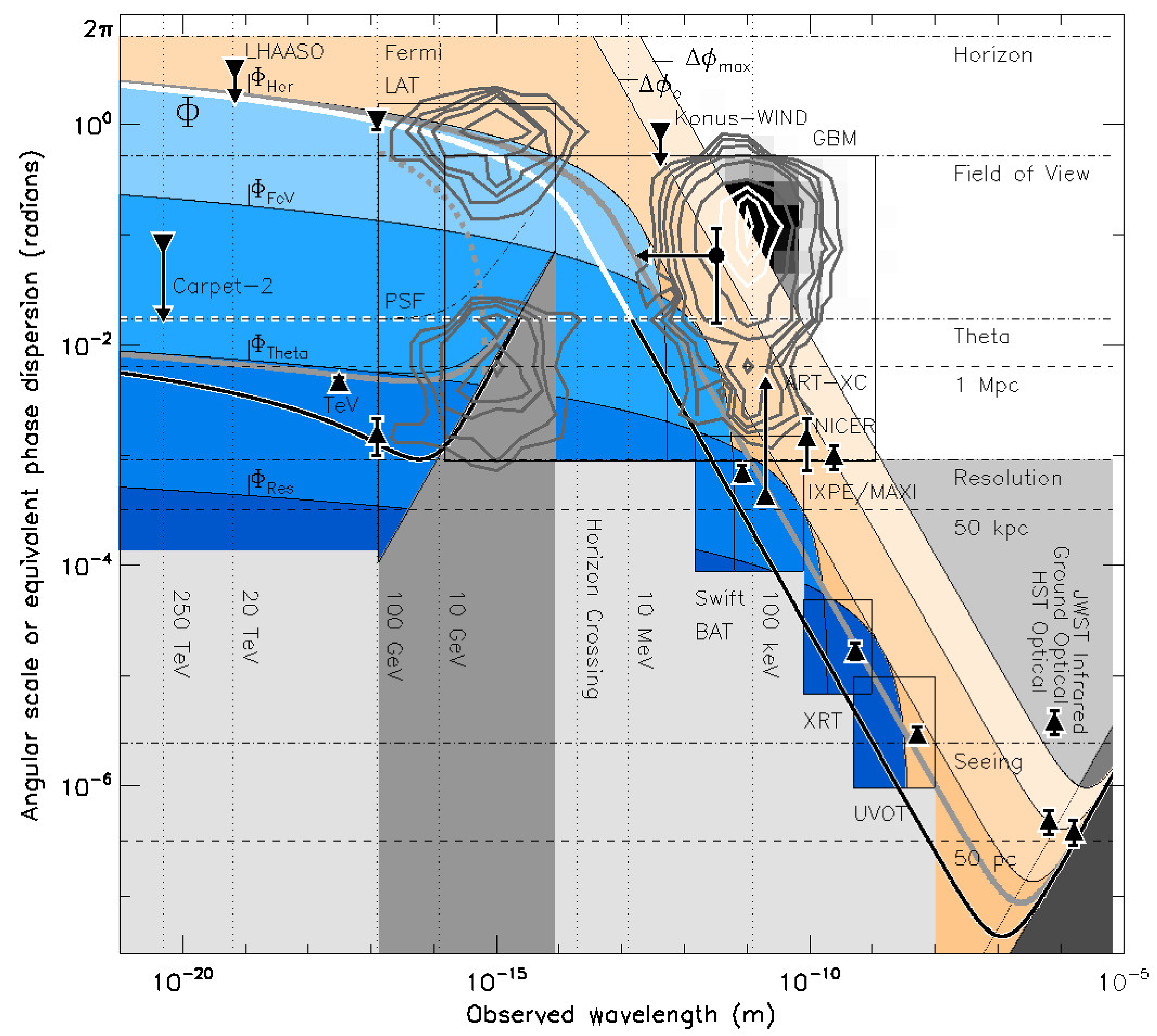}
\caption{Same as Figure~\ref{figure_model}, now with GRB measurements including GRB221009A; reproduced from \cite{Steinbring2023}. Density contours are of all available, archival GRB data from {\it Fermi} LAT (roll-angles and resolutions) or GBM (error-radii) shown in 10\% increments, white above 50\% for GBM. Note how the broadest case of a real instrumental PSF (white curve; equation 4) nicely matches the roll angle/zenith angle for the {\it Fermi}-LAT instrument (shown for $\alpha=0.650$, $z=0.151$, and black-on-white where below Theta; grey-curve: same, $z=1.41$) or the blueward edge of the error radius for the GBM ($\Phi_{\rm FoV}$). The lower limit where blurring must necessarily rise above the nominal resolution ($\Phi_{\rm Res}$) then occurs at $\alpha=0.735$, also shown as a black-on-white curve, where it is less than Theta. In between, scaling from an upper limit of $\Phi_{\rm Hor}$ to $\Phi_{\rm Theta}$, the last term in equation 4 would follow the ratio $1+2\pi/(1.22\times\theta)$, where $\theta$ is $1^\circ$. And so, the half-way point for $\alpha$ (i.e., just scaled instead by the ratio of the resolution to the horizon) happens to coincide with $\alpha=0.667$, which is the holographic value favored by QG models; the associated horizon-crossing angle is also indicated. This limiting behaviour - that an instrument records only the ensemble of wavelength-dependent scattering cases more than its resolution-limit and less than its FoV - is something like the effect of seeing from the ground, and leads to the smooth scaling for the average size of the bluest $\gamma$-ray sources (plotted as a dashed gray curve for the average redshift $z=1.41$). This regime will be scrutinized more carefully in the discussion to follow.}\label{figure_wide}
\end{figure}

This object triggered the {\it Fermi} space observatory Gamma-ray Burst Monitor (GBM) within its FoV of $35^\circ$ \citep{Veres2023}, which located it on the sky to within a error-radius of $3.71^\circ$ (90\% confidence) at peak energy 375 keV, and also with the Large Area Telescope \citep[LAT;][]{Bissaldi2023}. Those instrumental entendues are outlined in Figure~\ref{figure_wide}: the latter has a resolution of $5^\circ$ at 30 MeV or 1.5 arcmin at 60 GeV, and together with GBM has found over 3390 GRBs (median-$z$: 1.41; highest-$z$: 4.61) from 100 MeV to 100 GeV since launch in 2008; contour plots of all localization data available from the High-Energy Astrophysics Science Research Archive: https://heasarc.gsfc.nasa.gov/ (as accessed on 1 November 2022) plotted at the peak detected wavelength, as scaled by $1/(2\pi{\rm c}\hbar)$. Although GRB-monitoring instruments like LAT and GBM cannot resolve those sources, the LAT does provide a measure of how far the telescope must slew - called a roll angle - to centre the GRB within the instrument FoV, which is something less than the zenith angle to the celestial pole.

\vspace{3 mm}

Thus, once triggered, finding GRB221009A with LAT already restricts blurring, as photons were detected at 100 GeV within 1 radian for either measure, including at 397.7 GeV; so not all could have been scattered to the horizon. And, for the first time at least one higher-energy photon can be inferred by Carpet-2 via cosmic-ray air-shower angle-of-arrival (AoA) at 251 TeV. This instrument has an all-sky FoV (essentially $2\pi$) and resolution of $4.7^\circ$ (90\% confidence), with minimum set by the angular distance to the optical transient (down-pointing arrow in Figure~\ref{figure_wide}). This sets a critical angle near $1^\circ$, because if foam-induced blurring is present at this wavelength, it likewise demands here (and at all longer wavelengths) some wavefronts phase-dispersed less. That benchmark phase-error angle also happens to be near the mid-point of {\it Fermi} LAT/GBM resolutions, which is nominally the mean PSF of those instruments. For LAT it has been measured via images of a large sample of AGNs. Other relevant data are also shown in Figure~\ref{figure_wide}, including {\it Swift} satellite detection with the Burst-Alert Telescope (BAT) X-Ray Telescope (XRT), and Ultra-violet Optical Telescope (UVOT); black symbols indicate the positional accuracy, both roll-angle/zenith-angle or error-radius (above Theta) and localization/resolution (below Theta). The most-stringent of these in each energy range are plotted: at 18 TeV by the Large High Altitude Air Shower Observatory (LHAASO), allowing only a broad sky localization much like LAT roll-angle; and the space-based instruments Konus-{\it WIND} (KW) and {\it Mikhail Pavlinsky} Astronomical Roentgen Telescope X-ray Concentrator (ART-XC), the Neutron star Interior Composition Explorer (NICER) and Monitor of All-sky X-ray Imager (MAXI), the Imaging X-ray Polarimetry Explorer (IXPE) along with imaging from HST, the {\it James Webb Space Telescope} (JWST) and a redshift obtained from ground-based optical telescope spectroscopy. Thus, having located the object on the sky to within a degree in higher-energy $\gamma$-rays than seen before, and then later being able to identify its host galaxy, new restrictions can be placed on blurring attributable to spacetime foam.

\vspace{3 mm}

It is important to point out that LAT does not function like an optical telescope; it is a pair-production imaging instrument, sensitive to $\gamma$-rays through resultant electron/positron tracks. Scintillometers covering the array are used to reject background events, and a separate calorimeter recovers the incident $\gamma$-ray energy. However, the detected area on the sky of sensed $\gamma$-rays for a particular source does sample over what constitutes its resolution limit (although much larger than its diffraction limit) and an averaged fit of those (at a given energy) can be considered the instrumental PSF of the telescope. This has been measured, post-launch, with a large sample of AGNs, which gives the scaling relation $${\rm PSF}\propto\sqrt{[(C_0 E/100)^{-\rho}]^2 + C_1^2},$$where $E$ is energy in MeV; $C_0=3.5^\circ$ and $C_1=0.15^\circ$, and $\rho=0.8$ is the power-law slope \citep{Ackermann2013}. In Figure~\ref{figure_wide} this is plotted as a dashed curve, scaled to Theta.

\subsection{Seeing a Halo Around the Highest-Energy Quantum-Foam Blurred Gamma-Ray Sources}

\begin{figure}
\plotonesmall{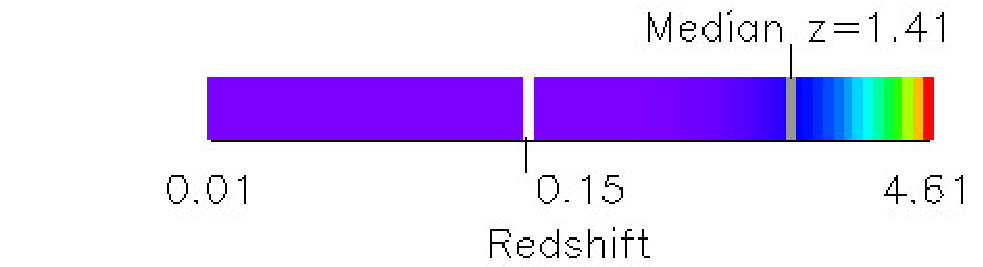}\\
\plotonemedium{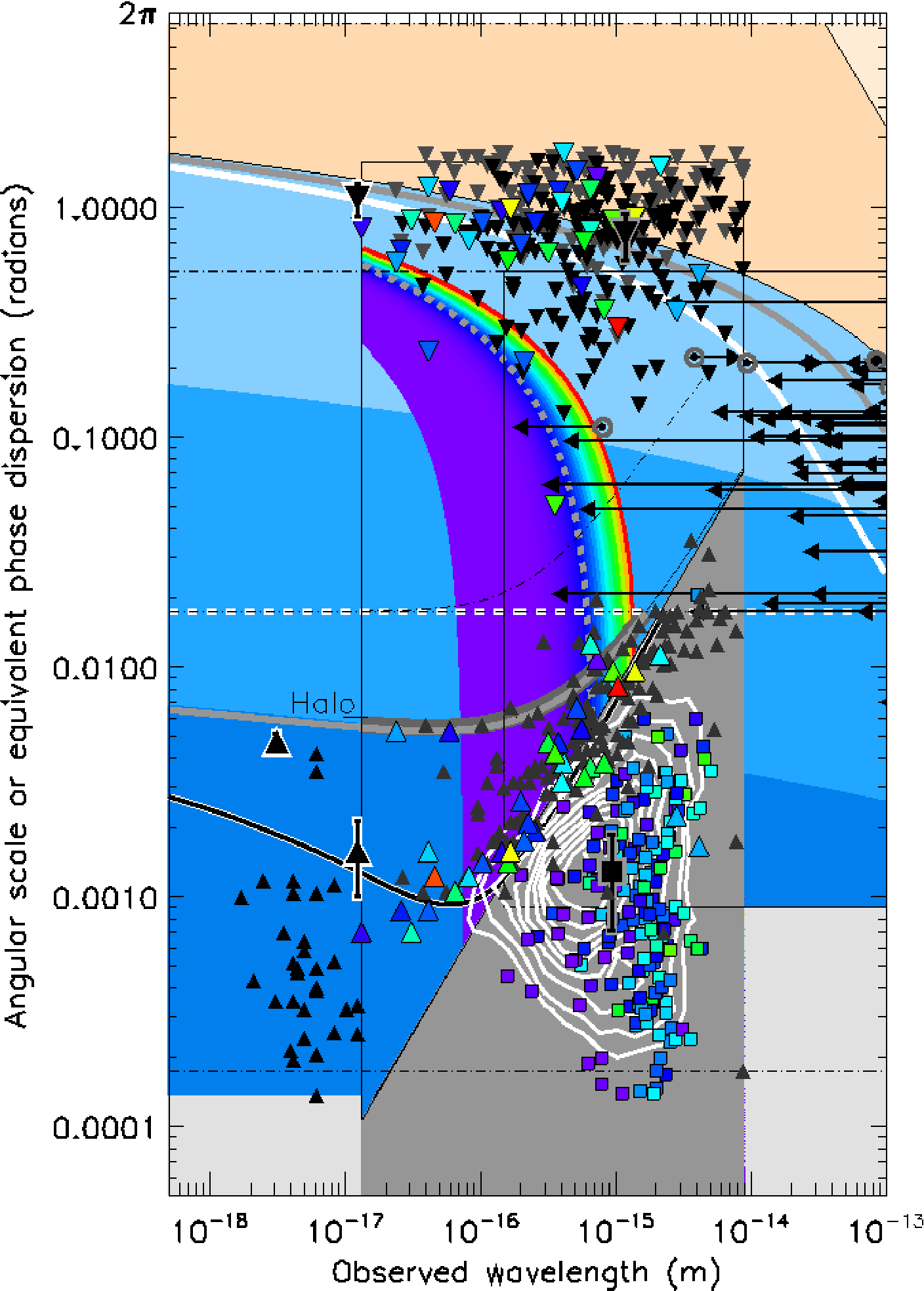}
\caption{As in Figure~\ref{figure_wide}, but further restricted to $\gamma$-ray wavelengths to show the highest-energy LAT photons detected. The down-pointing triangles are all archival roll-angles (black) and zenith angles (grey) available from the {\it Fermi} Archive as of 1 November 2022; up-pointing triangles are their corresponding resolutions. Some left-pointing triangles indicate the highest-energy detections from GBM; grey circles indicate the mean energy. At best PSF-sharpness, the open squares are the cataloged {\it Fermi} LAT resolutions measured for AGNs (each within a narrow energy range) and colour-coded by its redshift, where known; the PSF scaling behaviour described in Figure~\ref{figure_wide} is shown colour-coded in the same way. White density contours outline the full sample in 10\% increments. To the left, at higher energies than ever detected for AGNs, are TeV sources (black, up-pointing triangles) from the TeVCat catalog (complete as accessed on 1 November 2022 from http://tevcat.uchicago.edu). Model curves are as described before, but here light-grey: $\alpha=0.650$, $z=1.41$; black-on-white: $\alpha=0.735$, $z=0.151$. The most remarkable aspect of this plot is the agreement of the best-resolved highest-energy LAT GRBs with the dark-grey curve labelled ``Halo," as this would be the expected result, limited by quantum-foam for $\alpha=0.667$.}\label{figure_gamma}
\end{figure}

How well the expected upper-limit of blurring matches the mean distribution of roll-angles (including GRB221009A) is further illustrated in Figure~\ref{figure_gamma}; seen at upper-left, for LAT data by themselves. Below that, most remarkable is that the model of foam-induced-blurring plus LAT instrument resolution limit (thick grey curve labelled ``Halo") fits the locus of poorest-resolved GRBs well. And agreement nicely continues to where they turn off in approaching angle Theta at lowest LAT energies. In other words: this explains the ``tailing off" of largest image sizes at the longest wavelengths detectable with LAT, despite what naively might be expected to worsen with instrumental resolution (the slope of the dark-grey shaded region). That upper edge of GRB datapoints also closely matches the upper extent of AGN-size samples from LAT; taken from the Fourth pointsource catalogue (Abdollahi et al. 2020; 4050 samples, mean-$z$: 0.967), shown as white density contours (1\%, 5\%, 10\%, 25\%, 50\%, 75\%, 90\%; and, where available, filled squares color-coded by redshift using scale shown above, and from which the telescope instrumental PSF is reported. Equation 4 anticipates this behaviour, because if these are true point sources of a given redshift blurred only by foam, the last term in that equation allows that each can accumulate any image size {\it smaller} than Theta, down to the resolution limit - even though that may be less than the mean resolution at that energy; LAT images are not limited by diffraction itself. This instead implies that foam-induced blurring sets that lower limit: a resolution floor indicated here by the lower dot-dashed horizontal line. Notice that for the highest energies detectable with LAT (shortest wavelengths) the weakest possible $\Phi_{\rm Theta}$ (black on white curve, $\alpha=0.735$, $z=0.151$) is consistent with localization of GRB221009A. And in between, the average redshift (dashed curve) agrees with the mean energies of all LAT and GBM sources as well, that is, either the down- and left-pointing arrows for any of those sources could lie anywhere within the sensitivty of LAT and GBM, but blurred by holographic-QG foam ($\alpha=0.667$) none of those should stray blueward of this smooth, redshift-scaled demarcation (colour-coded in the same way as the symbols). That does seem to be the case, judging that one-sigma from the mean angular localization of the sample is a fair estimate of uncertainty in these reported data, accommodating two possible outliers (one dark blue and one green down-pointing triangle) in the LAT GRB roll-angle sample.

\section{Summary and Conclusions}\label{conclusions}

The PSF for holographic quantum-foam-induced blurring, $\Phi$, which depends on the telescope field of regard and energy range, has been presented. The classical analogy of a Plinko board is used to help illustrate that, although most photons may only be weakly affected by foam, it is the overall mean width Theta of the distribution in observed phase errors that ultimately matters. This is likewise analogous to how the image quality of a ground-based optical/near-infrared telescope gets smeared out by the uncorrected random refractive-index fluctuations induced by atmospheric turbulence along the beam, resulting in long exposures of pointsources being spread into a seeing disk.

\vspace{3 mm}

Accounting for realistic entendues of telescopes is able to explain both the broad multi-degree angular spread in localization of GRBs on the sky in high-energy $\gamma$-rays and X-rays, and how these can later still be identified at sub-arcesecond scales within their source galaxy via optical/near-infrared spectroscopy. That is key to understanding what appears to be happening: averaging over the senstivity ranges of real $\gamma$-ray telescopes (as imaging systems do by their nature) accounts for the consequences of some photons falling outside the telescope FoV (relevant to explaining the high-energy trend in roll-angle measurements) and some within the resolution and energy-sensitivity of an instrument; the fit is particularly good for data obtained with {\it Fermi} LAT, especially if $\alpha=0.667$, and suggests foam-induced blurring could be responsible for the resolution-limiting behavour of its PSF, which is of direct interest to $\gamma$-ray astronomy more generally.

\vspace{3 mm}

The recent GRB 221009A is special, as it was the brightest one ever observed, and visible across a huge wavelength range. Importantly, its highest-energy photons would be the most-scattered by quantum-foam, possibly with dispersions of many degrees.  Holographic foam-induced blurring sets a mean angular size $\Phi_{\rm Theta}$ of any GRB to be close to $1^\circ$ near 250 TeV, the highest energy localization of GRB221009A. Interpreted that way, GRB221009A provided the first clear observational evidence that spacetime is not smooth, although this observation remains consistent with all previous imaging of galaxies, AGNs, and all other available GRB localizations known.

\section*{Acknowledgments}

The author gratefully acknowledges discussions of Planck-scale effects with Richard Lieu, Eric Perlman, and especially Jack Ng, who also preceeded my talk at IARD 2024 with an excellent review of holographic quantum foam, regarding possible tests and observational expectations based on theory.


\begin{thebibliography}{}

\bibitem[Abdollahi et al.(2020)]{Abdollahi2020} Abdollahi, S., Acero, F., Ackermann, M., Ajello, M.  2020, Astrophs. Journal Supp. Series, 247 33
\bibitem[Ackermann et al.(2013)]{Ackermann2013} Ackermann, M., Ajello, M., Allafort, A., et al. 2013c, Astrophys. Journal, 765, 54
\bibitem[Bissaldi et al.(2023)]{Bissaldi2023} Bissaldi, E., Omodei, N., Kerr, M. et al. 2023, GRB Coordinates Network, Circular Service, 32637
\bibitem[Amelino-Camelia(2013)]{Amelino-Camelia2013} Amelino-Camelia, G. 2013, Living Reviews in Relativity, Volume 16, article number 5
\bibitem[Carlip(2023)]{Carlip2023} Carlip, S. 2023, Rep. Prog. Phys. 86 066001
\bibitem[Christiansen et al.(2011)]{Christiansen2011} Christiansen, W.A., Ng, Y.J., Floyd, D.J.E., \& Perlman, E.S. 2009, Physical Review D, 83, 84003
\bibitem[Lieu \& Hillman(2003)]{Lieu2003} Lieu, R. \& Hillman, L.W. 2003, Astrophys. Journal Lett., 585, L77 
\bibitem[Ng, Christiansen, \& van Dam(2003)]{Ng2003} Ng, Y.J., Christiansen, W.A., \& van Dam, H. 2003, Astrophs. Journal Lett., 591, L87
\bibitem[Ng \& Perlman(2022)]{Ng2022} Ng, Y.J. \& Perlman, E. 2022, Universe, 8, 382
\bibitem[Perlman et al.(2011)]{Perlman2011} Perlman, E.S., Ng, Y.J., Floyd, D.J.E., \& Christiansen, W.A. 2011, Aston. and Astrophys., 535, L9
\bibitem[Perlman et al.(2015)]{Perlman2015} Perlman, E.S., Rappaport, S.A., Christiansen, W.A., Ng, Y.J., DeVore, J., \& Pooley, D. 2015, Astrophys. Journal, 805, 10
\bibitem[Ragazzoni, Turatto, \& Gaessler(2003)]{Ragazzoni2003} Ragazzoni, R., Turatto, M., \& Gaessler, W. 2003, Astrophys. Journal Lett., 587, L1
\bibitem[Steinbring(2007)]{Steinbring2007} Steinbring, E. 2007, Astrophys. Journal, 655, 714
\bibitem[Steinbring(2015)]{Steinbring2015} Steinbring, E. 2015, Astrophys. Journal, 802, 38
\bibitem[Steinbring(2016)]{Steinbring2016} Steinbring, E. 2016, IAU Symposium Conf. Series, 319, 54
\bibitem[Steinbring(2023)]{Steinbring2023} Steinbring, E. 2023, Galaxies, 11(6), 115
\bibitem[Steinbring(2023)]{Steinbring2023b} Steinbring, E. 2023, Astrophysics Source Code Library, record ascl:2311.003
\bibitem[Tamburini et al.(2011)]{Tamburini2011} Tamburini, F., Cuofano, C., Della Valle, M., \& Gilmozzi, R. 2011, Astron. and Astrophys., 533, A71
\bibitem[Veres et al.(2023)]{Veres2023} Veres, P., Burns, E., Bissaldi, E., Lesage, S. et al. 2023, GRB Coordinates Network, Circular Service, 32636
\bibitem[Wheeler(1957)]{Wheeler1957} Wheeler, J.A. 1957, Annals of Physics, 2, 604

\end{thebibliography}
\end{document}